\documentclass{aa}  
\usepackage{graphicx}
\usepackage{txfonts}
\usepackage{amsmath}
\usepackage{xcolor}
\usepackage{multirow}
\usepackage[colorlinks=true, linkcolor=blue, urlcolor=blue, citecolor=blue,]{hyperref}
\bibpunct{(}{)}{;}{a}{}{,} 
\newcommand\scalemath[2]{\scalebox{#1}{\mbox{\ensuremath{\displaystyle #2}}}}
\newcommand*\chem[1]{\ensuremath{\mathrm{#1}}}

\begin{document} 

   \title{Magnetic fields in molecular clouds: Limitations of the analysis of Zeeman observations}

   \author{R. Brauer\inst{1},
          S. Wolf\inst{1},
          S. Reissl\inst{2},
          \and
          F. Ober\inst{1}
          }
          
   \authorrunning{R. Brauer et al.}

   \institute{University of Kiel, Institute of Theoretical Physics and Astrophysics,
              Leibnizstrasse 15, 24118 Kiel, Germany\\
              \email{[rbrauer;wolf]@astrophysik.uni-kiel.de}
           \and
               University of Heidelberg, Institute of Theoretical Astrophysics,
               Albert-Ueberle-Str. 2 U04, 69120 Heidelberg, Germany\\
               \email{reissl@uni-heidelberg.de}
	      }


 
  \abstract
   {Observations of Zeeman split spectral lines represent an important approach to derive the structure and strength of magnetic fields in molecular clouds. In contrast to the uncertainty of the spectral line observation itself, the uncertainty of the analysis method to derive the magnetic field strength from these observations is not been well characterized so far.
   }
   {We investigate the impact of several physical quantities on the uncertainty of the analysis method, which is used to derive the line-of-sight (LOS) magnetic field strength from Zeeman split spectral lines. These quantities are the density, temperature, velocity, and the magnetic field strength.
   }
   {We simulate the Zeeman splitting of the 1665~MHz OH line with the 3D radiative transfer (RT) extension ZRAD. This extension is based on the line RT code Mol3D \citep{ober_tracing_2015} and has been developed for the POLArized RadIation Simulator POLARIS \citep{reissl_radiative_2016}.
   }
   {Observations of the OH Zeeman effect in typical molecular clouds are not significantly affected by the uncertainty of the analysis method. However, some observations obtained a magnetic field strength of more than ${\sim}300~\mathrm{\mu G}$, which may result in an uncertainty of the analysis method of ${>}10\%$. We derived an approximation to quantify the range of parameters in which the analysis method works sufficiently accurate and provide factors to convert our results to other spectral lines and species as well. We applied these conversion factors to CN and found that observations of the CN Zeeman effect in typical molecular clouds are neither significantly affected by the uncertainty of the analysis method. In addition, we found that the density has almost no impact on the uncertainty of the analysis method, unless it reaches values higher than those typically found in molecular clouds ($n_\mathrm{H}\gg10^{7}~\mathrm{cm^{-3}}$). Furthermore, the uncertainty of the analysis method increases, if both the gas velocity and the magnetic field show significant variations along the line-of-sight. However, this increase should be small in Zeeman observations of most molecular clouds considering typical velocities of ${\sim}1~\mathrm{km/s}$.
   }  
   {}
 
   \keywords{ISM: clouds --
             Line: profiles --
             Magnetic fields --
             Polarization --
             Radiative transfer
             }

  \maketitle

\section{Introduction}\label{introduction}
The impact of magnetic fields on the formation of stars and planets is a matter of ongoing discussions \citep[e.g.,][]{matthews_magnetic_2002, pudritz_role_2014, seifried_impact_2015}. For instance, a comprehensive understanding of the relative importance of magnetic fields and turbulence in driving the formation of molecular cores has not been achieved so far \citep{crutcher_self-consistent_2010}. Magnetic fields are also expected to influence the shape of cloud fragments and the coupling between the gas and dust phase in molecular clouds \citep{henning_measurements_2001}.

Direct and indirect measurement methods are available to measure the structure and strength of magnetic fields in star-forming regions. Indirect measurements are performed by observing the polarized emission of elongated dust grains that align with their longer axis perpendicular to the  magnetic field lines \citep[e.g.,][]{bertrang_large-scale_2014, reissl_tracing_2014, brauer_origins_2016}. Based on the dispersions of the polarization direction and the line-of-sight (LOS) velocity, the so-called Chandrasekhar-Fermi method is used to calculate the magnetic field strength in the plane-of-sky \citep{chandrasekhar_magnetic_1953}. The uncertainty of this method is about ${\sim}20\%$ to ${\sim}200\%$ and depends on the spatial resolution and how well the velocity dispersion is known \citep{crutcher_observations_2004, cho_technique_2016}. On the other hand, direct measurements are performed by observing spectral lines that are split up by the Zeeman effect \citep{crutcher_magnetic_2012}. The uncertainty of the magnetic field strength derived from these observations range from ${\sim}5\%$ to a few $100\%$ \citep{crutcher_magnetic_2010}. With the sensitivity and spectral resolution of currently operating instruments, only the LOS component of the magnetic field strength can be obtained from Zeeman observations \citep{crutcher_oh_1993}. However, the total magnetic field strength can be estimated by using a Bayesian analysis \citep[see][]{crutcher_magnetic_2010} or by combining direct and indirect, i.e. complementary measurement methods \citep{heiles_magnetic_2012}.

Observations of Zeeman split CN, OH and HI lines have been used various times in the past to study the magnetic field in molecular clouds \citep{crutcher_magnetic_1983, crutcher_oh_1993, crutcher_detection_1999, heiles_millennium_2004, crutcher_magnetic_2010, crutcher_magnetic_2012, crutcher_magnetic_2014}. From such observations, the magnetic field strength in the LOS direction is usually derived following the approach by \cite{crutcher_oh_1993}. However, the limitations and uncertainties of this analysis method have not been well characterized so far. Therefore, we investigate the influence of several physical quantities on the uncertainty of this analysis method. These quantities are the density, temperature, velocity, and the magnetic field strength in the molecular cloud.

To achieve this goal, we perform radiative transfer simulations that consider the Zeeman splitting of the 1665~MHz OH line. For this, we use the RT code POLARIS \citep{reissl_radiative_2016} extended by our Zeeman splitting RT extension ZRAD, which is based on the line RT code Mol3D \citep{ober_tracing_2015}. With this code, we are able to solve the radiative transfer equation for Zeeman split OH lines by using the output of MHD simulations or analytical models.

In this study, we start with a description of the radiative transfer code that we use for our study (Sect. \ref{ZRAD}). Subsequently, we introduce the mentioned analysis method and the characteristics of the 1665~MHz OH line (Sect. \ref{model_description}). Our results are presented in Sect. \ref{results}, where we also describe our approach to investigate the uncertainty of the analysis method. After that, we investigate the influence of the density, temperature, turbulence and magnetic field strength on the analysis method, which is followed by the impact of variations in the velocity field. In Sect. \ref{other_zeeman}, we discuss how our results can be applied in the case of other transitions of OH and other species. Our conclusions are summarized in Sect. \ref{conclusions}.

\section{Radiative transfer}\label{ZRAD}
For this study, we apply the three-dimensional continuum RT code POLARIS \citep{reissl_radiative_2016}. It solves the RT problem self-consistently on the basis of the Monte Carlo method and allows one to consider magnetic fields as well as various dust grain alignment mechanisms. We extended POLARIS by the capability to consider the RT in spectral lines by implementing the line RT algorithm from the line RT code Mol3D \citep{ober_tracing_2015}. In addition, we developed the Zeeman splitting extension ZRAD that takes the Zeeman splitting and polarization of spectral lines into account. The implementation of the Zeeman splitting is based on the works of \cite{landi_deglinnocenti_malip_1976}, \cite{schadee_zeeman_1978}, \cite{rees_stokes_1989}, and \cite{larsson_treatment_2014}. For each considered Zeeman split spectral line, ZRAD requires the following pre-calculated quantities:
\begin{itemize}
 \item Energy levels and transitions taken from the Leiden Atomic and Molecular Database \citep[LAMDA,][]{schoier_atomic_2005},
 \item Land\'{e} factors of the involved energy levels,
 \item Line strengths of the allowed transitions between Zeeman sublevels, and
 \item Radius of the involved gas species.
\end{itemize}
For the line shape, ZRAD includes natural, collisional, and Doppler broadening mechanisms (line shape: Voigt profile) as well as the magneto-optic effect \citep[line shape: Faraday-Voigt profile,][]{larsson_treatment_2014}. The Voigt and Faraday-Voigt profiles are obtained from the real and imaginary part of the Faddeeva function, respectively \citep{wells_rapid_1999}. In ZRAD, a fast and precise solution of the Faddeeva function is realized with the Faddeeva package \citep{faddeeva_package}. 

\section{Model description}\label{model_description}
\subsection{Analysis method}\label{analysis_method}

The radiation of Zeeman split spectral lines can be described with the radiative transfer equation in the following form \citep{larsson_treatment_2014}:
\begin{align}
 \frac{\mathrm{d}\vec{I}_\nu}{\mathrm{d}s}&=-\mathbf{K}_{\nu}\left(\vec{I}_\nu-\vec{S}_\nu\right),\label{eq:rad_trans}\\
 \mathbf{K}_{\nu} &= \frac{n}{2} S_0 \sum_{M^\prime,M^{\prime\prime}}\left[S_{M^\prime,M^{\prime\prime}}F\left(\nu^\prime,a\right)\mathbf{A}_{M^\prime,M^{\prime\prime}}\right].\label{eq:prop}
\end{align}
Here, $M$ is the total angular or atomic momentum quantum number projected on the magnetic field vector $\vec{B}$. The quantity $S_0$ is the line strength of the spectral line without Zeeman splitting and $S_{M^\prime,M^{\prime\prime}}$ are the line strengths of the transitions between Zeeman sublevels. Single prime denotes the upper level, double prime denotes the lower level of the line transition. The quantity $n$ is the gas number density and $F\left(\nu^\prime,a\right)$ is the Voigt profile line shape function where:
\begin{align}
 \nu^\prime&=\frac{\nu_0+\Delta\nu_z-\nu}{\Delta\nu_\mathrm{D}},\ \mathrm{and}\\
 a&=\frac{\gamma}{4\pi\Delta\nu_\mathrm{D}}.
\end{align}
Here, $\nu$ is the frequency, $\nu_0$ is the frequency of the line peak, $\Delta\nu_z$ is the frequency shift due to Zeeman splitting, $\Delta\nu_\mathrm{D}$ is the Doppler broadening width, and $\gamma$ is a combination of the natural and collisional broadening width. The calculation of the broadening widths are described in detail by \cite{line_broadening}.

Only three transitions from $M^\prime$ to $M^{\prime\prime}$ are allowed. The transitions with $\Delta M=\pm1$ are called $\sigma_\pm$-transitions, whereby the transition with $\Delta M=0$ is called $\pi$-transition. The polarization rotation matrix $\mathbf{A}_{M^\prime,M^{\prime\prime}}$ depends on the type of transition as follows \citep{landi_deglinnocenti_malip_1976, rees_stokes_1989, larsson_treatment_2014}:
\begin{align}
 \scalemath{0.9}{\mathbf{A}_\mathrm{\sigma_\pm}}&=
 \scalemath{0.85}{\begin{pmatrix}
  1+\cos^2\theta          & \cos(2\eta)\sin^2\theta & \sin(2\eta)\sin^2\theta & \mp2\cos\theta\\
  \cos(2\eta)\sin^2\theta & 1+\cos^2\theta          & 0                       & 0 \\
  \sin(2\eta)\sin^2\theta & 0                       & 1+\cos^2\theta          & 0 \\
  \mp2\cos\theta          & 0                       & 0                       & 1+\cos^2\theta\\
 \end{pmatrix}},\ \mathrm{and} \label{eq:prop_sigma}\\
\scalemath{0.9}{\mathbf{A}_\mathrm{\pi}}&=
 \scalemath{0.83}{\begin{pmatrix}
  \sin^2\theta             & -\cos(2\eta)\sin^2\theta & -\sin(2\eta)\sin^2\theta & 0\\
  -\cos(2\eta)\sin^2\theta & \sin^2\theta             & 0                        & 0 \\
  -\sin(2\eta)\sin^2\theta & 0                        & \sin^2\theta             & 0 \\
  0                        & 0                        & 0                        & \sin^2\theta\\
 \end{pmatrix}}. \label{eq:prop_pi}
\end{align}
Here, $\theta$ is the angle between the magnetic field $\vec{B}$ and the LOS direction $\vec{R}$. The quantity $\eta$ is the clockwise angle between the vertical polarization direction and the projection of the magnetic field $\vec{B}$ on the normal plane of the LOS direction $\vec{R}$.

\begin{figure*}
 \centering
 \resizebox{\hsize}{!}{\includegraphics[width=\hsize, page=1]{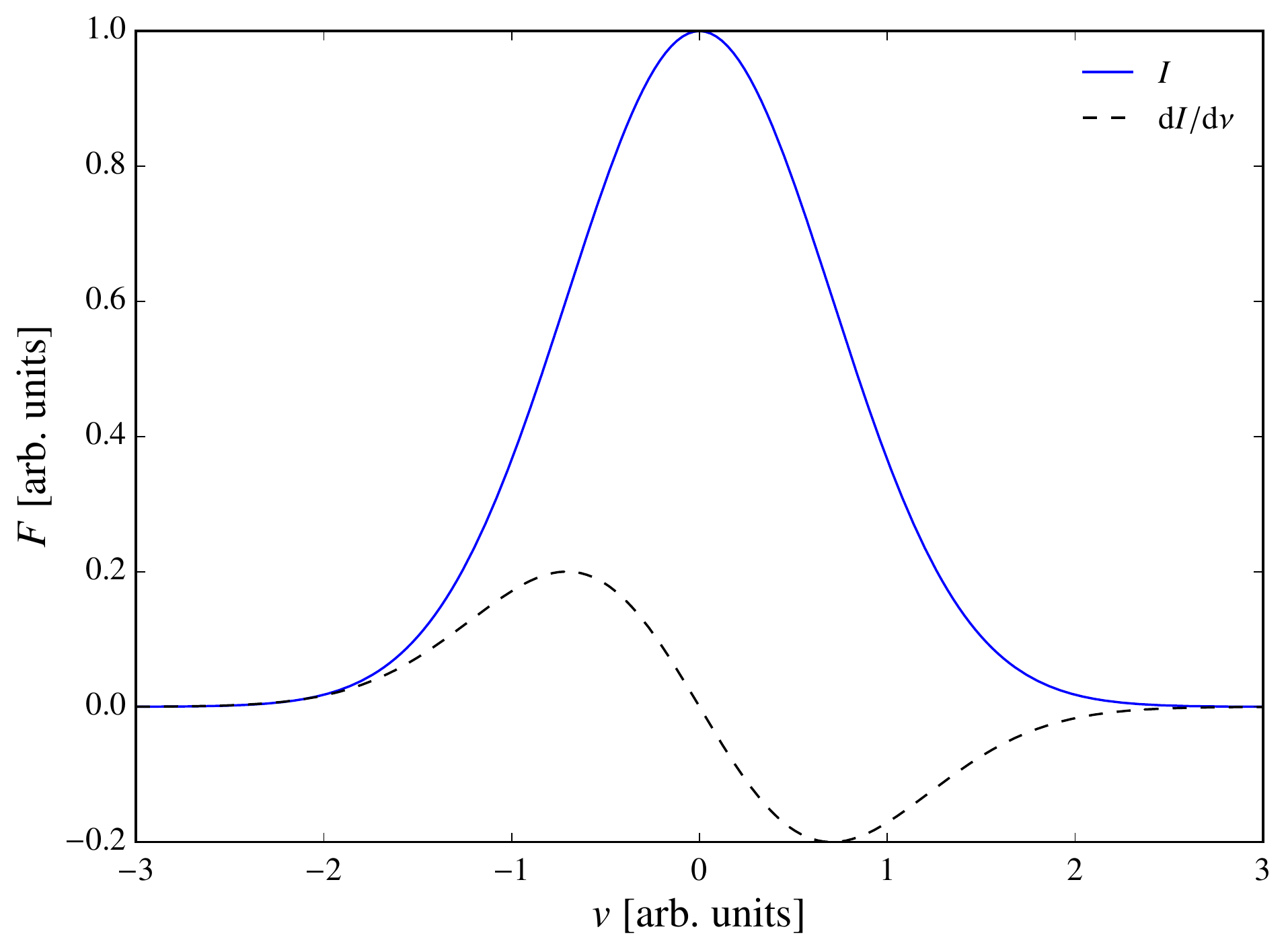} \quad
                       \includegraphics[width=\hsize, page=3]{figures/derive_mag_field_spectrum_2_new.pdf}}\\
 \resizebox{\hsize}{!}{\includegraphics[width=\hsize, page=2]{figures/derive_mag_field_spectrum_2_new.pdf} \quad
                       \includegraphics[width=\hsize, page=4]{figures/derive_mag_field_spectrum_2_new.pdf}}
 \caption{Schematic illustration of the analysis method to estimate the LOS magnetic field strength. The Zeeman shift is either negligible compared to the spectral line width (left) or in the same order of magnitude (right). In the lower left image, the ratio of $\mathrm{d}I/\mathrm{d}\nu$ to $V$ is proportional to the magnetic field strength in LOS direction $B\cos\theta$. The estimation of the LOS magnetic field strength from the spectral line in the lower right image varies significantly from the present LOS magnetic field strength in the model.}
 \label{fig:derive_mag_field_spectrum}
\end{figure*}

By combining Eqs. \ref{eq:rad_trans} to \ref{eq:prop_pi}, we obtain descriptions for the intensity $I$ and circular polarization $V$ similar to the expressions derived by \cite{crutcher_oh_1993}:
\begin{align}
 I=&F(\nu_0+\Delta\nu_z-\nu,a)\cdot(1+\cos^2\theta) 
 + 2F(\nu_0-\nu,a)\cdot\sin^2\theta\nonumber\\ \quad
 + &F(\nu_0-\Delta\nu_z-\nu,a)\cdot(1+\cos^2\theta), \\
 V=&\left[2F(\nu_0+\Delta\nu_z-\nu,a) - 2F(\nu_0-\Delta\nu_z-\nu,a)\right]\cos\theta. \label{eq:mag_field_0}
\end{align}
If we assume a line width $\Delta\nu$ which is significantly larger than the Zeeman shift $\Delta\nu_z$, we can derive the circular polarization as follows (see Appendix \ref{appendix_1} for details):
\begin{equation}
 V=\left(\frac{\mathrm{d}I}{\mathrm{d}\nu}\right)\Delta\nu_z\cos\theta, \label{eq:mag_field}
\end{equation}
whereby the frequency shift $\Delta\nu_z$ owing to Zeeman splitting can be calculated with:
\begin{equation}
  \Delta\nu_z=\frac{B\mu_\mathrm{b}}{h}(g^\prime M^\prime-g^{\prime\prime} M^{\prime\prime}).\label{eq:Zeeman_split}
\end{equation}
Here, $\mu_\mathrm{b}$ is the Bohr magneton and $g$ the Land\'{e} factor of the corresponding energy level. Following Eqs. \ref{eq:mag_field} and \ref{eq:Zeeman_split}, the magnetic field strength in the LOS direction can be obtained by \emph{fitting the derivative of the intensity profile to the circular polarization profile of a Zeeman split spectral line}. For illustration, Fig. \ref{fig:derive_mag_field_spectrum} shows this procedure for a spectral line that has a sufficiently low (left) or too large (right) Zeeman shift to estimate the LOS magnetic field strength with Eqs. \ref{eq:mag_field} and \ref{eq:Zeeman_split}.

\subsection{1665~MHz transition of OH}\label{1665MHz}
In this study, we perform our simulations for the $1665~\mathrm{MHz}$ OH line. In addition, we provide a discussion about the application of our results to observations of other Zeeman split spectral lines in Sect. \ref{other_zeeman}. The Zeeman splitting of the $1665~\mathrm{MHz}$ OH line is caused by the quantized orientation of the total atomic angular momentum on the magnetic field direction. The Land\'{e} factors of the energy levels involved in the $1665~\mathrm{MHz}$ OH line can be derived as follows \citep{radford_microwave_1961}:
\begin{align}
 g_\mathrm{F}&=g_\mathrm{J}\frac{F(F+1)+J(J+1)-K(K+1)}{2F(F+1)},\\
 g_\mathrm{J}&=\frac{1}{J(J+1)}\left\{\frac{3}{2}+\frac{2\left(J-\frac{1}{2}\right)\left(J+\frac{3}{2}\right)-\frac{3}{2}\lambda_\mathrm{OH}+3}{\left[4\left(J+\frac{1}{2}\right)^2+\lambda_\mathrm{OH}\left(\lambda_\mathrm{OH}-4\right)\right]^\frac{1}{2}}\right\}.
\end{align}
Here, $\lambda_\mathrm{OH}$ is a molecule-specific constant ($\lambda_\mathrm{OH}=-7.5$). The quantities $K$, $J$, and $F$ are the atomic, total orbital, and total atomic angular momentum quantum numbers (1665~MHz OH line: $K=\frac{1}{2}, J=\frac{3}{2}, F=1$). The relative line strengths of transitions between Zeeman sublevels of the $1665~\mathrm{MHz}$ OH line can be calculated as follows \citep{larsson_treatment_2014}:
\begin{align}
 \Delta M_F&=0:\qquad S_{M^\prime,M^{\prime\prime}}=\frac{3M_F^2}{F(F+1)(2F+1)}\\
 \Delta M_F&=\pm1:\qquad S_{M^\prime,M^{\prime\prime}}=\frac{3(F\mp M_F)(F+1\pm M_F)}{4F(F+1)(2F+1)}
\end{align}
Here, $M_F$ is the total atomic angular momentum quantum number $F$ projected on the magnetic field vector $\vec{B}$.

\section{Results}\label{results}
\subsection{Approach}\label{approach}
The analysis method is based on the assumption that the Zeeman shift is small in comparison to the line width. By altering the line shape, several quantities are expected to influence the validity of this assumption. These quantities are the temperature and turbulence (Doppler broadening), the density (collisional broadening), and the magnetic field strength. We investigate which range of values of these quantities allows a reliable estimation of the LOS magnetic field strength.
 
Since the gas temperature and turbulence influence the line width in the same way, we use the full width at half maximum (FWHM) of the Doppler broadening line width $\Delta \mathrm{v}^\mathrm{fwhm}_\mathrm{D}$ to consider both quantities at once:
\begin{equation}
 \Delta \mathrm{v}^\mathrm{fwhm}_\mathrm{D} = 2 \sqrt{\ln(2)} \sqrt{\frac{2 N_\mathrm{A} k_B T_\mathrm{OH}}{10^{-3}\cdot m_\mathrm{OH}} + \mathrm{v}^2_\mathrm{turb}}. \label{eq:fwhm_doppler}
\end{equation}
Here, $T_\mathrm{OH}$ is the gas temperature of the OH molecules, $m_\mathrm{OH}$ is the molar mass of OH in $[\mathrm{g/mol}]$, $N_\mathrm{A}$ is the Avogadro constant, $k_B$ is the Boltzmann constant, and $\mathrm{v}_\mathrm{turb}$ is the turbulent velocity.

\begin{table*}[htpb]
 \centering
 \caption{Overview of model parameters used to investigate the uncertainty of the analysis method.}
 \label{tab:parameter_study}
 \renewcommand{\arraystretch}{1.2}
 \begin{tabular}{lll}
  \hline
  \hline
  Magnetic field strength along LOS & $B_\mathrm{LOS}$ & $[0.3,3000]~\mathrm{\mu G}$ \\
  Hydrogen number density  &  $n_\mathrm{H}$ & $[2\cdot10^{1},2\cdot10^{7}]~\mathrm{cm^{-3}}$ \\
  Doppler broadening width (FWHM) & $\Delta \mathrm{v}^\mathrm{fwhm}_\mathrm{D}$ & $[0.05,1.5]~\mathrm{km/s}$ \\
  Spectral resolution & $\Delta \mathrm{v}_\mathrm{res}$ & $0.03~\mathrm{km/s}$\\
  & $\Delta \nu_\mathrm{res}$ & $0.165~\mathrm{kHz}$\\
  Species & & OH  \\
  Transition & $\nu_0$ & $1665~\mathrm{MHz}$ \\
  Abundance & $\mathrm{OH}/\mathrm{H}$ & $10^{-7}$ \\
  \hline
 \end{tabular}
 \renewcommand{\arraystretch}{1}
\end{table*}

For the parameter space, we consider typical conditions in observed molecular clouds \citep{frerking_structure_1987, wilson_density_1997, minamidani_submillimeter_2008, crutcher_magnetic_2010, burkhart_observational_2015}. These are a hydrogen number density of  $9.1~\mathrm{cm^{-3}}$ to $1.3\cdot10^{7}~\mathrm{cm^{-3}}$, a gas temperature of $10~\mathrm{K}$ to $300~\mathrm{K}$, a turbulent velocity of $100~\mathrm{m/s}$ to $10~\mathrm{km/s}$, and a LOS magnetic field strength of $0.1~\mathrm{\mu G}$ to $3100~\mathrm{\mu G}$. Our parameter space (see Table \ref{tab:parameter_study}) approximately covers these conditions. Each quantity of the parameter space is realized with 128 logarithmically distributed values. We use an abundance of OH/H of $10^{-7}$ that is similar to values used in the works of \cite{crutcher_nonthermal_1979} and \cite{roberts_distribution_1995}: $\mathrm{OH}/\mathrm{H}=4\cdot10^{-8}$ and $\mathrm{OH}/\mathrm{H}=4\cdot10^{-7}$, respectively. We consider a spectral resolution of $\Delta \mathrm{v}_\mathrm{res}=0.03~\mathrm{km/s}$ that is similar to those of recent observations \citep[][$\Delta \mathrm{v}_\mathrm{res}=0.034~\mathrm{km/s}$ and $\Delta \mathrm{v}_\mathrm{res}=0.055~\mathrm{km/s}$ respectively]{troland_magnetic_2008, crutcher_testing_2009}.

Our model space is a cube. It is filled with optical thin gas of constant density. The level population of OH is calculated under the assumption of local thermodynamic equilibrium (LTE). However, the choice of the particular method for estimation of the level population is arbitrary, since the analysis method depends only on the ratio of the intensity $I$ to the circular polarization $V$, but not on the net flux.

To estimate the uncertainty of the analysis method, we compare the LOS magnetic field strength derived by the analysis method $B_\mathrm{LOS,\ derived}$ (Eq. \ref{eq:mag_field}, \ref{eq:Zeeman_split}; Sect \ref{analysis_method}) with the reference field strength $B_\mathrm{LOS}$. We calculate the reference field strength by averaging the LOS magnetic field strength weighted with the spectral line intensity $I$ along the line-of-sight to the observer:
\begin{equation}
 B_\mathrm{LOS} = \frac{ \int_{0}^\mathrm{obs} \left(\vec{B}(s) \cdot \vec{e_\mathrm{LOS}}\right) I(s) \mathrm{d}s }{ \int_{0}^\mathrm{obs} I(s) \mathrm{d}s }
\end{equation}
Here, $\vec{B}(s)$ is the magnetic field strength, $\vec{e_\mathrm{LOS}}$ is the unit vector in the LOS direction, and $s$ is the path length along the line-of-sight. The relative difference between the derived and the reference LOS magnetic field strength is defined as follows:
\begin{equation}
 \frac{|\Delta B_\mathrm{LOS}|}{B_\mathrm{LOS}} = \frac{|B_\mathrm{LOS,\ derived} - B_\mathrm{LOS}|}{B_\mathrm{LOS}}.
\end{equation}
In this study, we refer to this quantity as the uncertainty of the analysis method and assume that a sufficient reliability is achieved if this uncertainty drops below $10\%$.

Alternatively, by using the same equations, quantities and assumptions presented in this section and Sect. \ref{model_description}, it is possible to replicate the results of the following Section with a simpler approach by using multiple Gaussians and comparing the Zeeman shift with the line width. However, the radiative transfer simulations allow us to consider more complex scenarios with variable physical quantities such as density, velocity field, and magnetic field strength (see Sect. \ref{velocity_field}).

\subsection{Density, temperature, turbulence, and magnetic field strength}\label{dens_temp_mag}

\begin{figure}
 \centering
 \includegraphics[width=\hsize]{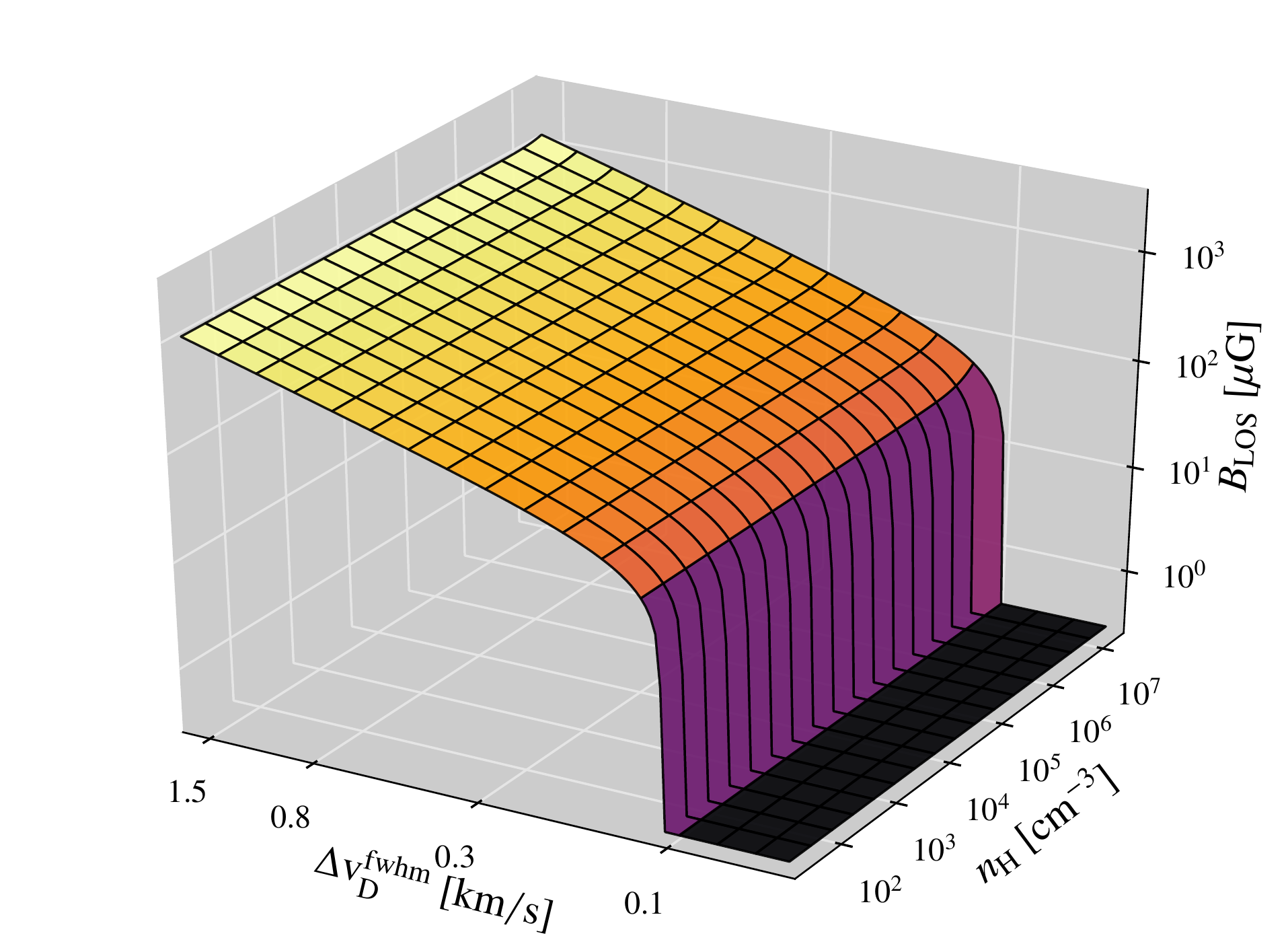}
 \caption{Contour surface where the relative difference between the derived and the reference LOS magnetic field strength amounts to $10\%$. Each parameter combination below the surface corresponds to a lower relative difference.}
 \label{fig:parameter_study}
\end{figure}

The contour surface in Fig. \ref{fig:parameter_study} shows the LOS magnetic field strength at which the uncertainty of the analysis method is equal to $10\%$ within the considered parameter space. With increasing magnetic field strength, the uncertainty of the analysis method increases. Within typical values of the hydrogen number density, no significant impact of the density on the analysis method can be observed. Therefore, the Doppler effect is the dominant line broadening mechanism in molecular clouds ($\Delta\mathrm{v}\approx\Delta\mathrm{v}_\mathrm{D}$). Only if the density is very high ($n_\mathrm{H}>10^{7}~\mathrm{cm^{-3}}$ and thus larger than usually measured in molecular clouds), the LOS magnetic field strength at which the uncertainty of the analysis method reaches $10\%$ increases with increasing density.

\begin{figure*}
 \centering
 \resizebox{\hsize}{!}{\includegraphics[width=\hsize, page=4]{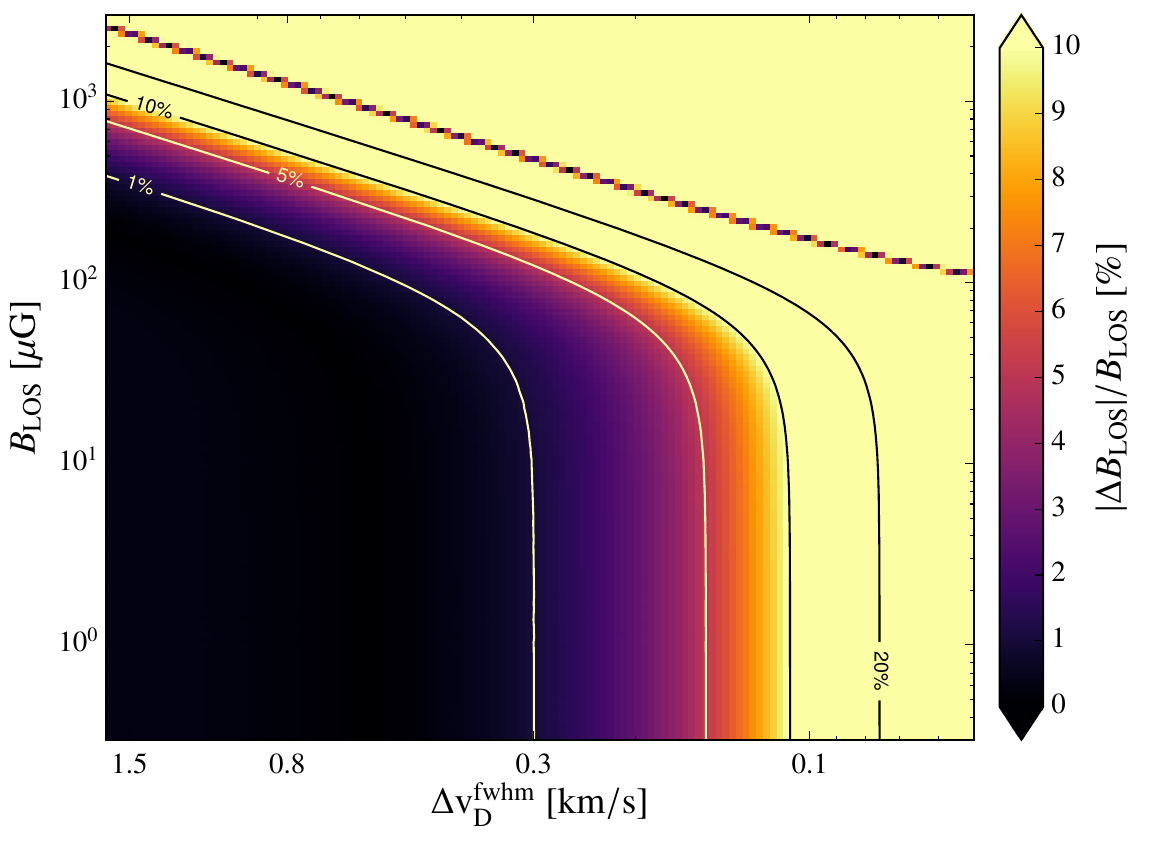} \quad
                       \includegraphics[width=\hsize, page=3]{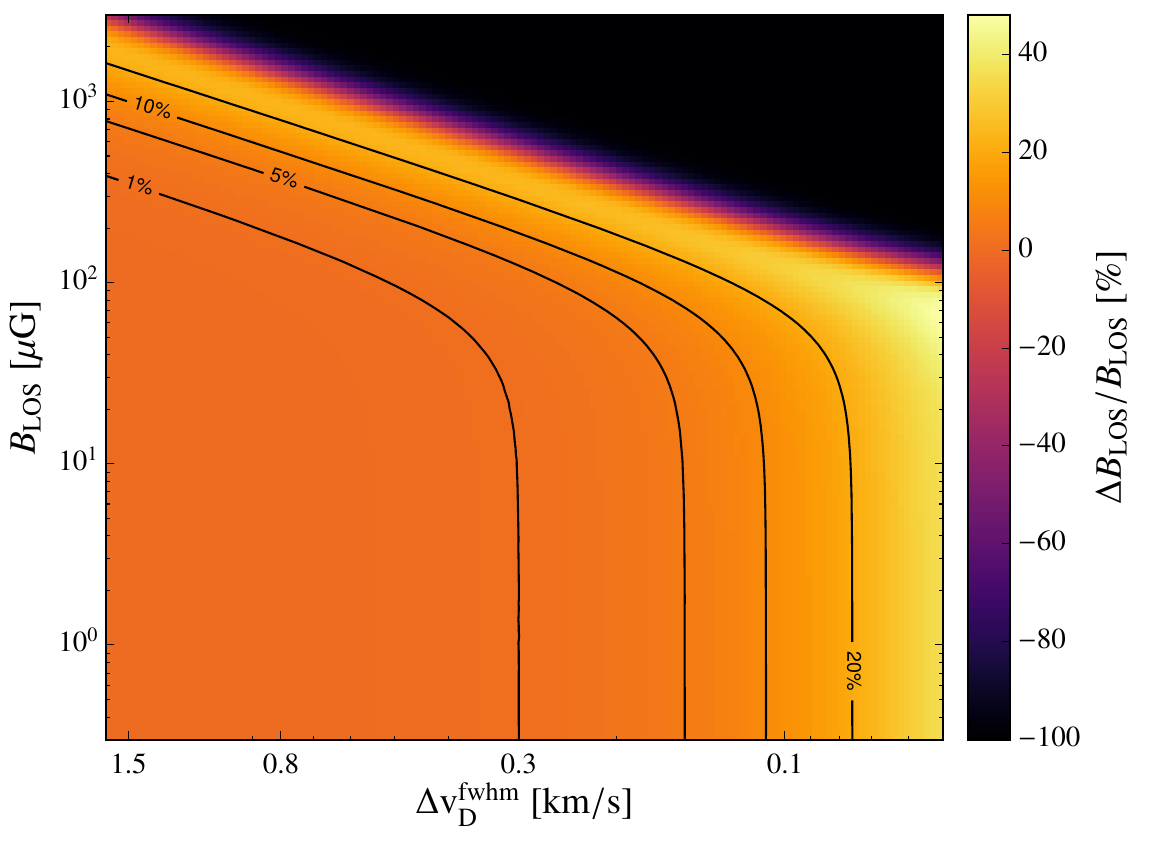}}
 \caption{Relative difference between the derived and the reference LOS magnetic field strength dependent on the LOS magnetic field strength $B_\mathrm{LOS}$ and the Doppler broadening width $\Delta \mathrm{v}^\mathrm{fwhm}_\mathrm{D}$ (hydrogen number density $n_\mathrm{H}=2\cdot10^{5}~\mathrm{cm^{-3}}$). The right image takes the sign of the relative difference into account. The contour lines correspond to relative differences of $1\%$, $5\%$, $10\%$, and $20\%$.}
 \label{fig:mag_to_temp_study}
\end{figure*}

Because of the weak dependence of the hydrogen number density on the uncertainty of this method, we now focus our study on the influence of the magnetic field strength and the Doppler broadening owing to temperature and turbulence (see Fig. \ref{fig:mag_to_temp_study}). With increasing line width, the assumption of the analysis method ($\Delta\nu\gg\Delta\nu_z$) is valid even in the case of larger Zeeman shifts, i.e. magnetic field strengths. In addition, spectral lines with a line width of less than ${\sim}0.1~\mathrm{km/s}$ always show an uncertainty of more than $10\%$. Due to the spectral resolution of $\Delta \mathrm{v}_\mathrm{res}=0.03~\mathrm{km/s}$, such lines are too narrow to be resolved sufficiently. 

\textbf{Local minimum:} In Fig. \ref{fig:mag_to_temp_study} (left), a narrow region of minimum relative difference between the derived and the reference LOS magnetic field strength can be seen (above the $20\%$ contour line). Here, the analysis method appears to allow one to obtain a very precise LOS magnetic field strength. However, this is not the case. Instead, the assumption of the analysis method is highly violated in this region. For illustration, Fig. \ref{fig:mag_to_temp_study} (right) shows the relative difference between the derived and the reference LOS magnetic field strength with their corresponding sign. In this narrow region, the relative difference changes its sign and, therefore, the uncertainty of the analysis method has to go through a minimum. Below this narrow region, the analysis method overestimates the LOS magnetic field strength and above, the analysis method underestimates the LOS magnetic field strength. As a result, observations of very high LOS magnetic field strengths (${>}1~\mathrm{mG}$) are likely to measure a significantly lower value than present in the cloud, whereby an observed LOS magnetic field strengths of some ${\sim}100~\mathrm{\mu G}$ could be up to ${\sim}50\%$ higher.

\textbf{Magnetic field strength $\perp$ LOS:}
The results shown in Figs. \ref{fig:parameter_study} and \ref{fig:mag_to_temp_study} consider a magnetic field with field vectors only pointing in the LOS direction. An additional magnetic field component perpendicular to the LOS direction adds the $\pi$-transition to the intensity profile that changes the superposed line shape (see Eq. \ref{eq:prop_pi}; Sect. \ref{analysis_method}). In addition, with increasing magnetic field strength perpendicular to the LOS direction, the Zeeman shift $\Delta\nu_z$ increases without measuring a higher LOS magnetic field strength. Therefore, we investigate the impact of a magnetic field component perpendicular to the LOS direction on our previous results. For this, we simulate the same cubic model and parameter space as shown in Table \ref{tab:parameter_study}, but add a magnetic field component perpendicular to the LOS direction with a strength of either three or ten times the LOS component.

\begin{figure*}
 \centering
 \resizebox{\hsize}{!}{\includegraphics[width=\hsize, page=4]{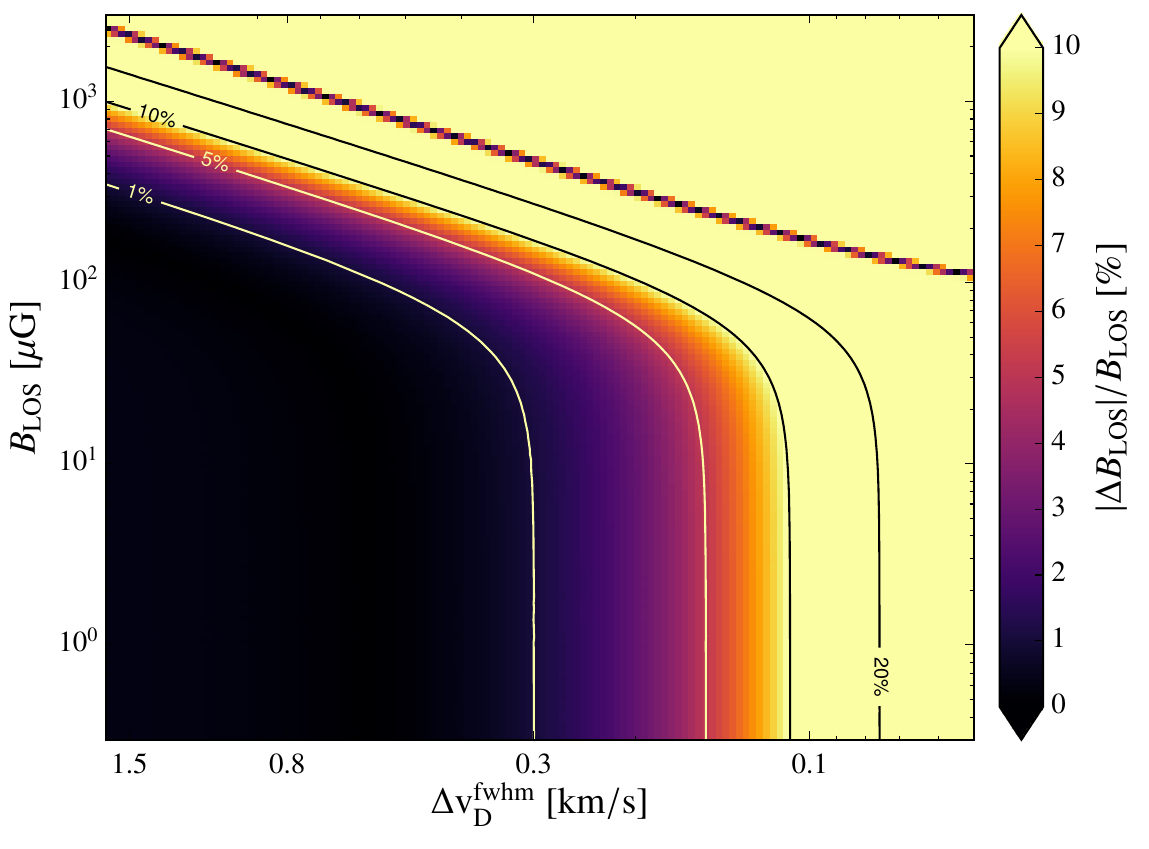} \quad
                       \includegraphics[width=\hsize, page=4]{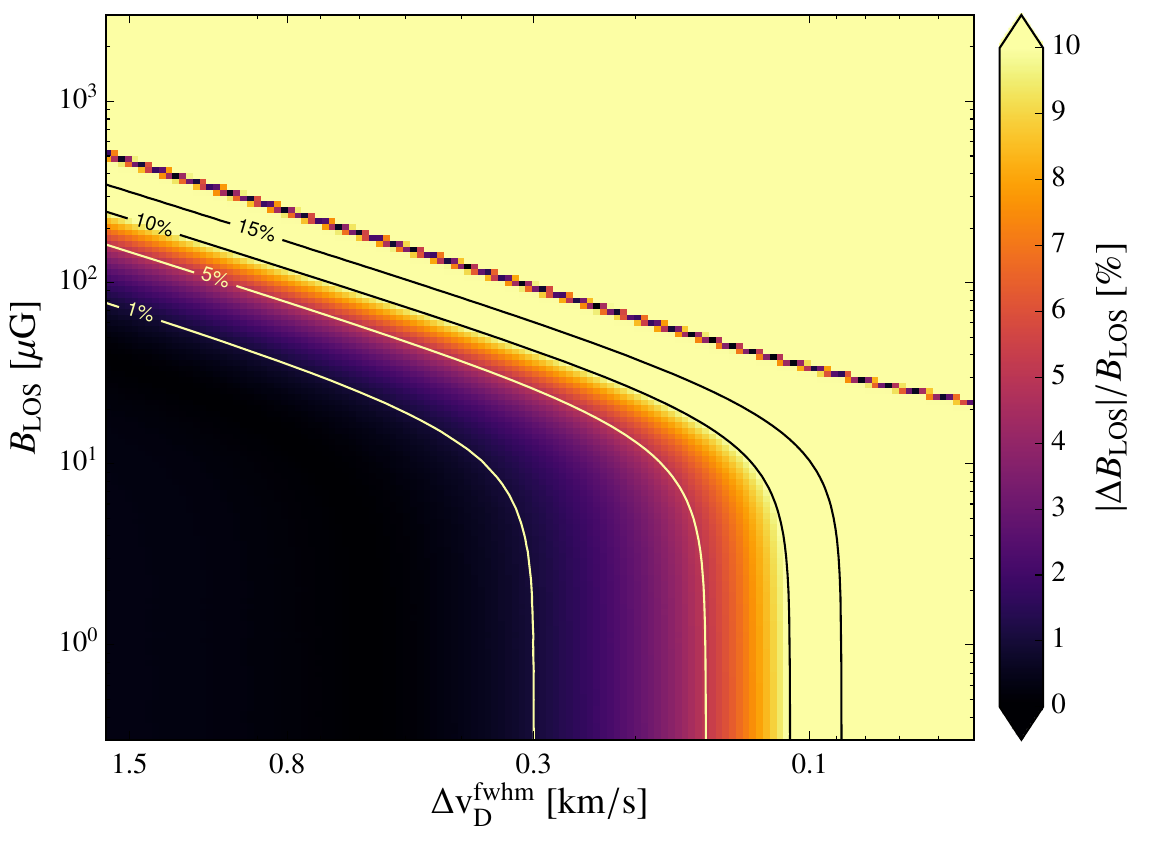}}
 \caption{Relative difference between the derived and the reference LOS magnetic field strength dependent on the LOS magnetic field strength $B_\mathrm{LOS}$ and the Doppler broadening width $\Delta \mathrm{v}^\mathrm{fwhm}_\mathrm{D}$ (hydrogen number density $n_\mathrm{H}=2\cdot10^{5}~\mathrm{cm^{-3}}$). The magnetic field strength perpendicular to the LOS is either three (left) or ten (right) times the LOS magnetic field strength. The contour lines correspond to relative differences of $1\%$, $5\%$, $10\%$, and $20\%$ (left) or $1\%$, $5\%$, $10\%$, and $15\%$ (right).}
 \label{fig:high_b_perp}
\end{figure*}

As expected, the uncertainty of the analysis method depends on the value of the Zeeman shift $\Delta\nu_z$ and therefore on the total magnetic field strength (see Fig. \ref{fig:high_b_perp}). Hence, an upper limit of the total magnetic field strength is necessary to constrain the uncertainty of the analysis method. \cite{crutcher_magnetic_1999} mentioned that the total magnetic field strength can be estimated by taking two times the mean LOS magnetic field strength. In this case, the impact of a magnetic field component perpendicular to the LOS direction is rather small.

We repeat the simulations shown in Fig. \ref{fig:high_b_perp} with total magnetic field strengths from $0.5~\mathrm{\mu G}$ to $100~\mathrm{mG}$ and orientations of the magnetic field vector relative to the line-of-sight from $0^\circ$ to $89^\circ$. By fitting $B^2_\mathrm{LOS}=a\cdot T^b - B^2_\perp$ to these results, we obtain an approximation to estimate the maximum LOS magnetic field strength, which can be derived with an uncertainty of less than ${\sim}10\%$:
\begin{equation}
\left(\frac{B_\mathrm{LOS}}{\mathrm{mG}}\right)^2 \lesssim 0.4 \cdot \left(\frac{\Delta \mathrm{v}^\mathrm{fwhm}_\mathrm{D}}{\mathrm{km/s}}\right)^2 - \left(\frac{B_\perp}{\mathrm{mG}}\right)^2 \label{eq:perp_thumb}
\end{equation}
Here, $B_\perp$ is the magnetic field strength perpendicular to the LOS direction and $\Delta \mathrm{v}^\mathrm{fwhm}_\mathrm{D}$ is the FWHM Doppler broadening width (see Eq. \ref{eq:fwhm_doppler}; Sect. \ref{approach}). This equation is applicable only for the Zeeman split $1665~\mathrm{MHz}$ OH line. However, we show in Sect. \ref{other_zeeman} how this equation can be adapted to other spectral lines as well.

\subsection{Velocity field}\label{velocity_field}
In our previous simulations, we neglected the impact of the possible motion of the gas, besides turbulent motion, on the analysis method. However, strong variations in the LOS component of the corresponding velocity field should significantly influence the shape of the spectral lines and, therefore, influence the analysis method. To investigate this, we simulate a Bonnor-Ebert sphere with a constant central density which can be written as follows \citep{kaminski_role_2014}:
\begin{equation}
  \rho(r)=
  \begin{cases}
    \hspace{0.1cm} \rho_0 \cdot R_0^{-2}, \quad \mathrm{if\ } r \leq R_0\\
    \hspace{0.1cm} \hspace{0.07cm}\rho_0 \cdot r^{-2}, \quad \mathrm{if\ } R_0 < r \leq R_\mathrm{out}\\
    \hspace{0.1cm} \hspace{0.9cm}0, \quad \mathrm{if\ } r > R_\mathrm{out}\\
  \end{cases}
  \label{eqn:density}
\end{equation}
Here, $\rho_0$ is a reference density used to achieve a given gas mass, $r$ is the radial distance to the center, and $R_\mathrm{0}$ is a truncation radius that defines the extent of the central region with constant density. We consider a total gas mass of $1~\mathrm{M_\odot}$ and an outer radius of $1.5\cdot10^4~\mathrm{AU}$ \citep[similar to Bok globule B335;][]{wolf_magnetic_2003}. The resulting hydrogen number density is in the same range as used in our previous simulations (see Table \ref{tab:parameter_study} and \ref{tab:parameter_velocity}). Therefore, the Doppler broadening width determines the line width, which we have chosen to be in agreement with observations \citep[see Table \ref{tab:parameter_velocity};][]{falgarone_cn_2008}. For the velocity field, we assume a collapse with a constant velocity of $1~\mathrm{km/s}$, which is in agreement with velocities found in other studies \citep{campbell_contraction_2016, wiles_scaled_2016}. For the magnetic field, we use an approximation found by \cite{crutcher_magnetic_2010}, which can be written as follows:
\begin{equation}
  B_\mathrm{LOS} = 10~\mathrm{\mu G} \left(\frac{n_\mathrm{H}}{300~\mathrm{cm^{-3}}}\right)^{0.65}.
\end{equation}
Here, $n_\mathrm{H}$ is the hydrogen number density and the magnetic field is oriented in the LOS direction.

\begin{figure}
  \centering
  \includegraphics[width=0.99\hsize, page=6]{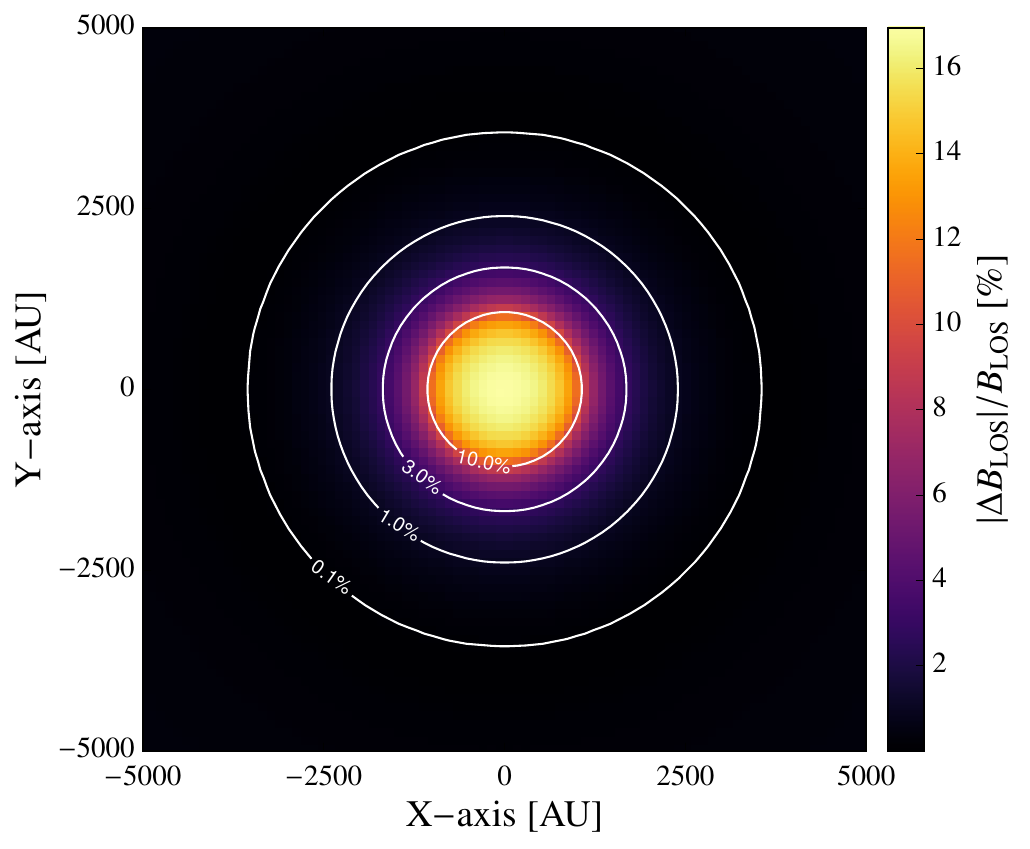}
  \caption{Relative difference between the derived and the reference LOS magnetic field strength of a spherical model with a Bonnor-Ebert sphere density structure and parameters summarized in Table \ref{tab:parameter_velocity}. The image shows the innermost $5000~\mathrm{AU}\times5000~\mathrm{AU}$. The velocity field is chosen to reproduce a collapsing molecular cloud. The contour lines correspond to relative differences of $0.1\%$, $1\%$, $3\%$, and $10\%$.}
  \label{fig:collapse}
\end{figure}

As illustrated in Fig. \ref{fig:collapse}, the uncertainty of the analysis method reaches values larger than $10\%$. This is caused by the velocity dependent frequency shift of the spectral lines that are related to different magnetic field strengths. If the frequency shift is comparable to the line width, multiple Zeeman split spectral lines are considered in the fitting process. As a consequence, it is not possible to exactly reconstruct the intensity averaged magnetic field strength, if both the gas velocity and the magnetic field show significant variations along the line-of-sight. However, this increase of the uncertainty of the analysis method should be small in Zeeman observations of most molecular clouds considering typical velocities in the order of magnitude of ${\sim}1~\mathrm{km/s}$.

\begin{table*}
 \centering
 \caption{Overview of model parameters used to investigate the influence of the motion of the gas, characterized by the gas velocity $\mathrm{v}_\mathrm{gas}$, on the analysis method.}
 \label{tab:parameter_velocity}
 \renewcommand{\arraystretch}{1.2}
 \begin{tabular}{lll}
  \hline
  \hline
  Magnetic field strength along LOS & $B_\mathrm{LOS}$ & $[60, 2000]~\mathrm{\mu G}$ \\
  Hydrogen number density  &  $n_\mathrm{H}$ & $[4\cdot10^{3}, 1\cdot10^{5}]~\mathrm{cm^{-3}}$ \\
  Doppler broadening width (FWHM) & $\Delta \mathrm{v}^\mathrm{fwhm}_\mathrm{D}$ & $1.5~\mathrm{km/s}$ \\
  Gas velocity & $\mathrm{v}_\mathrm{gas}$ & $2~\mathrm{km/s}$ \\
  Total gas mass & $M_\mathrm{gas}$ & $1~\mathrm{M_\odot}$ \\
  Outer radius  &  $R_\mathrm{out}$ & $1.5\cdot10^4~\mathrm{AU}$ \\
  Truncation radius & $R_\mathrm{0}$ & $10^3~\mathrm{AU}$ \\
  \hline
 \end{tabular}
 \renewcommand{\arraystretch}{1}
\end{table*}
 
\section{Zeeman splitting of other spectral lines}\label{other_zeeman}
In this study, we only consider the $1665~\mathrm{MHz}$ transition of OH. However, by taking particular coefficients into account, our results can be easily applied to other Zeeman split spectral lines as well. For another spectral line $i$ of species $j$, the Doppler broadening causes a different line width compared to OH due to the different molecular mass of the species (Eqs. \ref{eq:fwhm_doppler}; Sect. \ref{approach}). The following equation can be used to convert the Doppler broadening width of spectral line $i$ and species $j$ to those used for the $1665~\mathrm{MHz}$ transition of OH in Figs. \ref{fig:parameter_study} - \ref{fig:high_b_perp} and Eq. \ref{eq:perp_thumb} (Sect. \ref{results}):
\begin{align}
 \Delta \mathrm{v}^\mathrm{fwhm}_\mathrm{D,1665\mathrm{MHz}, OH} &= \sqrt{a_{j}}b_{i}\Delta \mathrm{v}^\mathrm{fwhm}_\mathrm{D,j} \nonumber\\
 &= \sqrt{\frac{17~\mathrm{g/mol}}{m_j}} \left(\frac{\nu_{i,0}}{1665~\mathrm{MHz}}\right) \Delta \mathrm{v}^\mathrm{fwhm}_\mathrm{D,j}. \label{eq:convert_d}
\end{align}
Here, $m_j$ is the molar mass of species $j$ and $\nu_{i,0}$ is the peak frequency of spectral line $i$. The Doppler broadening acts in units of the velocity and the Zeeman shift in units of the frequency. Therefore, a change in the spectral line peak frequency changes the line width compared to the Zeeman shift. In Eq. \ref{eq:convert_d}, this is considered by the coefficient $b_{i}$.  As a result of this equation, spectral lines with higher peak frequencies and species with a lower mass would provide a larger line width if compared to the Zeeman shift and therefore lower uncertainty of the analysis method. However, with decreasing Zeeman shift compared to the line width, the feasibility to detect the Zeeman effect decreases as well (see Eq. \ref{eq:mag_field}).

For another spectral line $i$ of species $j$, the Zeeman shift per magnetic field strength $\Delta\nu_z/B$ usually differs from the Zeeman shift of OH. By taking this into account, the magnetic field strength can be converted to the field strength used in Figs. \ref{fig:parameter_study} - \ref{fig:high_b_perp} and Eq. \ref{eq:perp_thumb} (Sect. \ref{results}):
\begin{equation}
 B_\mathrm{1665\mathrm{MHz}, OH}=c_{i,j} B_{i,j}. \label{eq:convert_b}
\end{equation}
Here, $c_{i,j}$ is the conversion factor of the chosen Zeeman split spectral line $i$ of species $j$ and can be calculated with the following equation (see Eq. \ref{eq:Zeeman_split}):
\begin{align}
 c_{i,j} &= \frac{1}{1.63~\mathrm{Hz/\mu G}}\frac{\Delta\nu_z}{B} \nonumber\\
 &= \frac{1}{1.63~\mathrm{Hz/\mu G}}\frac{\mu_\mathrm{b}}{h}(g_{i,j}^\prime M^\prime-g_{i,j}^{\prime\prime} M^{\prime\prime}).\label{eq:convert_c}
\end{align}

By combining Eq. \ref{eq:perp_thumb} from Sect. \ref{dens_temp_mag} with Eqs. \ref{eq:convert_d} and \ref{eq:convert_b}, the maximum LOS magnetic field strength, which can be derived with an uncertainty of less than ${\sim}10\%$ can be written as follows:
\begin{equation}
 \left(\frac{B_\mathrm{LOS}}{\mathrm{\mu G}}\right)^2
 \lesssim 0.4 \cdot
 \left(\frac{\sqrt{a_{j}} b_{i}}{c_{i,j}}\cdot \frac{\Delta \mathrm{v}^\mathrm{fwhm}_\mathrm{D}}{\mathrm{m/s}}\right)^2
 - \left(\frac{B_\perp}{\mathrm{\mu G}}\right)^2
 \label{eq:convert_all}
\end{equation}

In Table \ref{tab:conversion_factors}, we provide an overview of conversion factors for different spectral lines and species that can be used to perform Zeeman observations. We also summarize line widths and magnetic field strengths from observations to estimate if the analysis method provides a negligible uncertainty. It appears that observations of the Zeeman effect in typical molecular clouds are not significantly affected by the uncertainty of the analysis method. Even the spectral lines and species that provide no observations yet, should be usable up to magnetic field strengths typically observed in molecular clouds.

\begin{table*}[htpb]
 \centering
 \caption{Conversion factors and applicability of the analysis method for different spectral lines and species. The conversion factor $\frac{\sqrt{a} b}{c}$ and the maximum magnetic field strength up to which the analysis method can be used are calculated with Eqs. \ref{eq:convert_d}, \ref{eq:convert_c}, and \ref{eq:convert_all}.}
 \label{tab:conversion_factors}
 \renewcommand{\arraystretch}{1.2}
 \begin{tabular}{cccccc}
  \hline
  \hline
  \multirow{2}{*}{Species} & Peak frequency & \multirow{2}{*}{$\large\frac{\sqrt{a} b}{c}$} & \multicolumn{2}{c}{Typical values from observations} & Analysis method safely \\
  & $\nu_{0}$ [$\mathrm{MHz}$] & & Line width $\Delta\mathrm{v}$ [$\mathrm{km/s}$] & Field strength $B_\mathrm{LOS}$ [$\mathrm{\mu G}$] & usable up to $B$ [$\mathrm{\mu G}$]\\
  \hline
  OH$^{(1,3)}$ & $1665.4$ & $1$ & ${\sim}[0.5,15]$ & ${\sim}[10,3100]$ & ${<}300$ \\
  OH$^{(1,3)}$ & $1667.3$ & ${\sim}1.7$ & ${\sim}[0.5,15]$ & ${\sim}[10,3100]$ & ${<}500$ \\
  CN$^{(2)}$ & $113\,144.3$ & ${\sim}100$ & ${\sim}[1,4]$ & ${\sim}[10,1100]$ & ${<}60\,000$ \\
  CN$^{(2)}$ & $113\,170.9$ & ${\sim}680$ & ${\sim}[1,4]$ & ${\sim}[10,1100]$ & ${<}430\,000$ \\
  CN$^{(2)}$ & $113\,191.3$ & ${\sim}340$ & ${\sim}[1,4]$ & ${\sim}[10,1100]$ & ${<}220\,000$ \\
  CN$^{(2)}$ & $113\,488.4$ & ${\sim}100$ & ${\sim}[1,4]$ & ${\sim}[10,1100]$ & ${<}60\,000$ \\
  CN$^{(2)}$ & $113\,491.2$ & ${\sim}380$ & ${\sim}[1,4]$ & ${\sim}[10,1100]$ & ${<}240\,000$ \\
  CN$^{(2)}$ & $113\,499.7$ & ${\sim}340$ & ${\sim}[1,4]$ & ${\sim}[10,1100]$ & ${<}220\,000$ \\
  CN$^{(2)}$ & $113\,509.1$ & ${\sim}130$ & ${\sim}[1,4]$ & ${\sim}[10,1100]$ & ${<}80\,000$ \\
  SO$^{(1)}$ & $99\,299.9$ & ${\sim}110$ & $-$ & $-$ & $-$ \\
  SO$^{(1)}$ & $138\,178.6$ & ${\sim}200$ & $-$ & $-$ & $-$ \\
  SO$^{(1)}$ & $158\,971.8$ & ${\sim}180$ & $-$ & $-$ & $-$ \\
  SO$^{(1)}$ & $219\,949.9$ & ${\sim}510$ & $-$ & $-$ & $-$ \\
  SO$^{(1)}$ & $236\,452.3$ & ${\sim}160$ & $-$ & $-$ & $-$ \\
  \hline
 \end{tabular}
 \tablebib{(1)~\citet{bel_zeeman_1989}; (2) \citet{falgarone_cn_2008}; (3) \citet{crutcher_magnetic_1999}.}
 \renewcommand{\arraystretch}{1}
\end{table*}

\begin{figure}
 \centering
 \resizebox{\hsize}{!}{\includegraphics[width=\hsize, page=4]{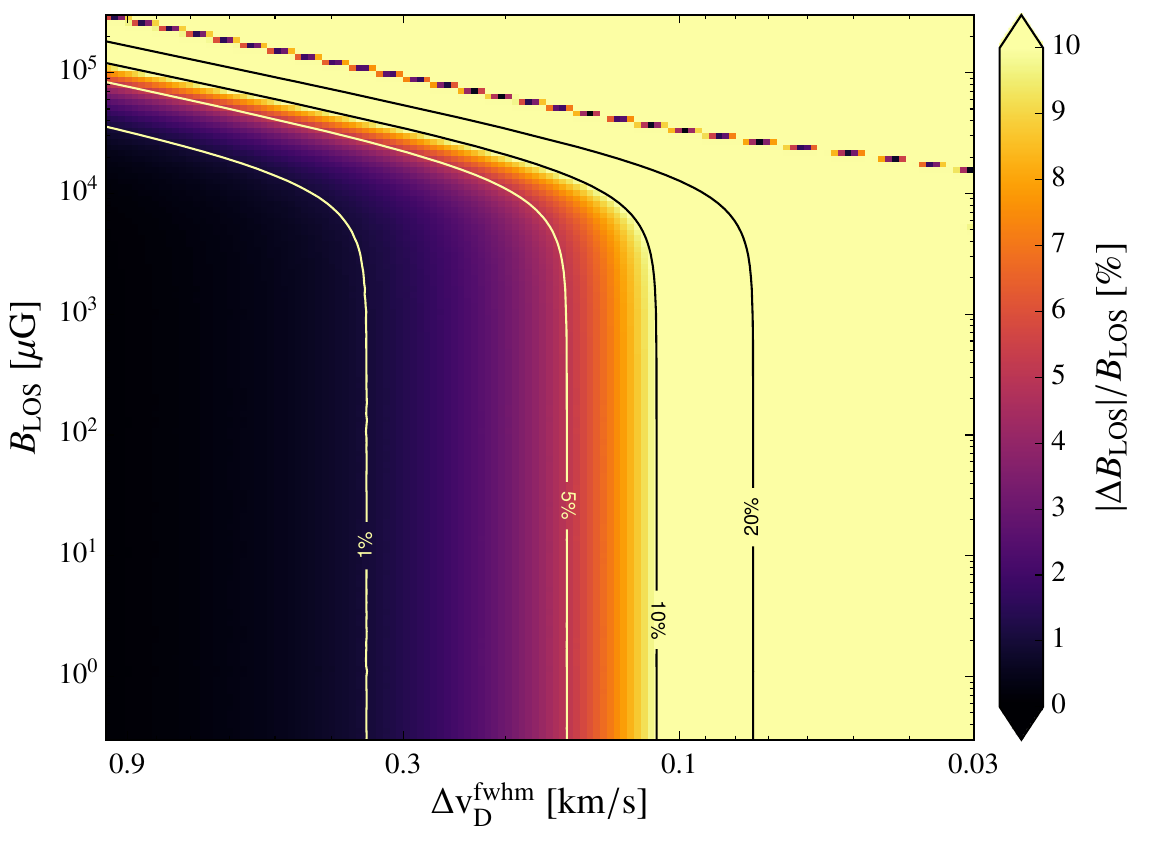}}
 \caption{Relative difference between the derived and the reference LOS magnetic field strength dependent on the LOS magnetic field strength $B_\mathrm{LOS}$ and the Doppler broadening width $\Delta \mathrm{v}^\mathrm{fwhm}_\mathrm{D}$ (hydrogen number density $n_\mathrm{H}=2\cdot10^{5}~\mathrm{cm^{-3}}$). The contour lines correspond to relative differences of $1\%$, $5\%$, $10\%$, and $20\%$.}
 \label{fig:mag_to_temp_study_SO}
\end{figure}

To check the validity of the conversion factors, we perform simulations with the $99.3~\mathrm{GHz}$ SO line instead of the $1665~\mathrm{MHz}$ OH line \citep[Zeeman splitting parameters from][]{clark_magnetic_1974, bel_zeeman_1989, larsson_treatment_2014}. We use the same parameters as shown in Table \ref{tab:parameter_study} of Sect. \ref{dens_temp_mag}. As expected by the conversion factor in Table \ref{tab:conversion_factors} and Eq. \ref{eq:convert_all}, the maximum LOS magnetic field strength, which can be derived with an uncertainty of less than ${\sim}10\%$ shifts by a factor of ${\sim}100$ to higher field strengths (see Fig. \ref{fig:mag_to_temp_study_SO}).

\section{Conclusions}\label{conclusions}
Observations of Zeeman split spectral lines in star-forming regions, i.e. molecular clouds, represent an established method to study the local magnetic field. We performed simulations with various parameter sets to estimate the value range in which the Zeeman analysis method is applicable for these observations. We found the following results:
\begin{enumerate}
 \item Observations of the OH Zeeman effect in typical molecular clouds are not significantly affected by the uncertainty of the analysis method. However, some observations obtained a magnetic field strength of more than ${\sim}300~\mathrm{\mu G}$, which may result in an uncertainty of the analysis method of ${>}10\%$.
 \item Besides OH, multiple species can be used to perform Zeeman observations (CN, HI, other promising species are \chem{C_2H}, \chem{SO}, \chem{C_2S}, \chem{C_4H}, and \chem{CH}; \citealt{crutcher_magnetic_2012}). We provided equations which allows one to apply our results to other Zeeman split spectral lines of various species as well (Eqs. \ref{eq:convert_d}, \ref{eq:convert_b}, and \ref{eq:convert_all}, Table \ref{tab:conversion_factors}; Sect. \ref{other_zeeman}). We applied these equations to CN and found that observations of the CN Zeeman effect in typical molecular clouds are neither significantly affected by the uncertainty of the analysis method.
 \item Within typical values for molecular clouds, the density has almost no impact on the uncertainty of the analysis method, unless it reaches values higher than those typically found in molecular clouds ($n_\mathrm{H}\gg10^{7}~\mathrm{cm^{-3}}$). As a consequence, the Doppler effect is the dominant broadening mechanism.
 \item The uncertainty of the analysis method increases, if both the gas velocity and the magnetic field show significant variations along the line-of-sight. However, this increase should be small in Zeeman observations of most molecular clouds considering typical velocities of ${\sim}1~\mathrm{km/s}$.
\end{enumerate}

 \begin{acknowledgements}
       Part of this work was supported by the German
       Deut\-sche For\-schungs\-ge\-mein\-schaft, DFG\/ project number WO 857/12-1.
 \end{acknowledgements}

\nocite{hunter_matplotlib:_2007}
\bibliographystyle{aa}
\bibliography{my_library,custom_bibtex}

\begin{thebibliography}{47}
\expandafter\ifx\csname natexlab\endcsname\relax\def\natexlab#1{#1}\fi

\bibitem[{Bel \& Leroy(1989)}]{bel_zeeman_1989}
Bel, N. \& Leroy, B. 1989, Astronomy and Astrophysics, 224

\bibitem[{Bertrang {et~al.}(2014)Bertrang, Wolf, \&
  Das}]{bertrang_large-scale_2014}
Bertrang, G., Wolf, S., \& Das, H.~S. 2014, Astronomy and Astrophysics, 565,
  A94

\bibitem[{Brauer {et~al.}(2016)Brauer, Wolf, \& Reissl}]{brauer_origins_2016}
Brauer, R., Wolf, S., \& Reissl, S. 2016, Astronomy \& Astrophysics, 588, A129

\bibitem[{Burkhart {et~al.}(2015)Burkhart, Collins, \&
  Lazarian}]{burkhart_observational_2015}
Burkhart, B., Collins, D.~C., \& Lazarian, A. 2015, The Astrophysical Journal,
  808, 48

\bibitem[{{Böttcher}(2012)}]{line_broadening}
{Böttcher}, M. 2012, {Line broadening mechanisms}

\bibitem[{Campbell {et~al.}(2016)Campbell, Friesen, Martin, Caselli, Kauffmann,
  \& Pineda}]{campbell_contraction_2016}
Campbell, J.~L., Friesen, R.~K., Martin, P.~G., {et~al.} 2016, The
  Astrophysical Journal, 819, 143

\bibitem[{Chandrasekhar \& Fermi(1953)}]{chandrasekhar_magnetic_1953}
Chandrasekhar, S. \& Fermi, E. 1953, The Astrophysical Journal, 118, 113

\bibitem[{Cho \& Yoo(2016)}]{cho_technique_2016}
Cho, J. \& Yoo, H. 2016, ArXiv e-prints, 1603, arXiv:1603.08537

\bibitem[{Clark \& Johnson(1974)}]{clark_magnetic_1974}
Clark, F.~O. \& Johnson, D.~R. 1974, The Astrophysical Journal, 191, L87

\bibitem[{Crutcher(2014)}]{crutcher_magnetic_2014}
Crutcher, R. 2014

\bibitem[{Crutcher(1979)}]{crutcher_nonthermal_1979}
Crutcher, R.~M. 1979, The Astrophysical Journal, 234, 881

\bibitem[{Crutcher(1999)}]{crutcher_magnetic_1999}
Crutcher, R.~M. 1999, The Astrophysical Journal, 520, 706

\bibitem[{Crutcher(2004)}]{crutcher_observations_2004}
Crutcher, R.~M. 2004, 123--132

\bibitem[{Crutcher(2012)}]{crutcher_magnetic_2012}
Crutcher, R.~M. 2012, Annual Review of Astronomy and Astrophysics, 50, 29

\bibitem[{Crutcher {et~al.}(2009)Crutcher, Hakobian, \&
  Troland}]{crutcher_testing_2009}
Crutcher, R.~M., Hakobian, N., \& Troland, T.~H. 2009, The Astrophysical
  Journal, 692, 844

\bibitem[{Crutcher {et~al.}(2010{\natexlab{a}})Crutcher, Hakobian, \&
  Troland}]{crutcher_self-consistent_2010}
Crutcher, R.~M., Hakobian, N., \& Troland, T.~H. 2010{\natexlab{a}}, Monthly
  Notices of the Royal Astronomical Society, 402, L64

\bibitem[{Crutcher \& Kazes(1983)}]{crutcher_magnetic_1983}
Crutcher, R.~M. \& Kazes, I. 1983, Astronomy and Astrophysics, 125, L23

\bibitem[{Crutcher {et~al.}(1993)Crutcher, Troland, Goodman, Heiles, Kazes, \&
  Myers}]{crutcher_oh_1993}
Crutcher, R.~M., Troland, T.~H., Goodman, A.~A., {et~al.} 1993, The
  Astrophysical Journal, 407, 175

\bibitem[{Crutcher {et~al.}(1999)Crutcher, Troland, Lazareff, Paubert, \&
  Kazès}]{crutcher_detection_1999}
Crutcher, R.~M., Troland, T.~H., Lazareff, B., Paubert, G., \& Kazès, I. 1999,
  The Astrophysical Journal Letters, 514, L121

\bibitem[{Crutcher {et~al.}(2010{\natexlab{b}})Crutcher, Wandelt, Heiles,
  Falgarone, \& Troland}]{crutcher_magnetic_2010}
Crutcher, R.~M., Wandelt, B., Heiles, C., Falgarone, E., \& Troland, T.~H.
  2010{\natexlab{b}}, The Astrophysical Journal, 725, 466

\bibitem[{Falgarone {et~al.}(2008)Falgarone, Troland, Crutcher, \&
  Paubert}]{falgarone_cn_2008}
Falgarone, E., Troland, T.~H., Crutcher, R.~M., \& Paubert, G. 2008, Astronomy
  and Astrophysics, 487, 247

\bibitem[{Frerking {et~al.}(1987)Frerking, Langer, \&
  Wilson}]{frerking_structure_1987}
Frerking, M.~A., Langer, W.~D., \& Wilson, R.~W. 1987, The Astrophysical
  Journal, 313, 320

\bibitem[{Heiles \& Haverkorn(2012)}]{heiles_magnetic_2012}
Heiles, C. \& Haverkorn, M. 2012, Space Science Reviews, 166, 293

\bibitem[{Heiles \& Troland(2004)}]{heiles_millennium_2004}
Heiles, C. \& Troland, T.~H. 2004, The Astrophysical Journal Supplement Series,
  151, 271

\bibitem[{Henning {et~al.}(2001)Henning, Wolf, Launhardt, \&
  Waters}]{henning_measurements_2001}
Henning, T., Wolf, S., Launhardt, R., \& Waters, R. 2001, The Astrophysical
  Journal, 561, 871

\bibitem[{Hunter(2007)}]{hunter_matplotlib:_2007}
Hunter, J.~D. 2007, Computing in Science and Engineering, 9, 90

\bibitem[{{Johnson}(2012)}]{faddeeva_package}
{Johnson}, S.~G. 2012, Faddeeva Package: complex error functions

\bibitem[{Kaminski {et~al.}(2014)Kaminski, Frank, Carroll, \&
  Myers}]{kaminski_role_2014}
Kaminski, E., Frank, A., Carroll, J., \& Myers, P. 2014, The Astrophysical
  Journal, 790, 70

\bibitem[{Landi~Degl'Innocenti(1976)}]{landi_deglinnocenti_malip_1976}
Landi~Degl'Innocenti, E. 1976, Astronomy and Astrophysics Supplement Series,
  25, 379

\bibitem[{Larsson {et~al.}(2014)Larsson, Buehler, Eriksson, \&
  Mendrok}]{larsson_treatment_2014}
Larsson, R., Buehler, S.~A., Eriksson, P., \& Mendrok, J. 2014, Journal of
  Quantitative Spectroscopy and Radiative Transfer, 133, 445

\bibitem[{Matthews \& Wilson(2002)}]{matthews_magnetic_2002}
Matthews, B.~C. \& Wilson, C.~D. 2002, The Astrophysical Journal, 574, 822

\bibitem[{Minamidani {et~al.}(2008)Minamidani, Mizuno, Mizuno, Kawamura,
  Onishi, Hasegawa, Tatematsu, Ikeda, Moriguchi, Yamaguchi, Ott, Wong, Muller,
  Pineda, Hughes, Staveley-Smith, Klein, Mizuno, Nikolić, Booth, Heikkilä,
  Nyman, Lerner, Garay, Kim, Fujishita, Kawase, Rubio, \&
  Fukui}]{minamidani_submillimeter_2008}
Minamidani, T., Mizuno, N., Mizuno, Y., {et~al.} 2008, The Astrophysical
  Journal Supplement Series, 175, 485

\bibitem[{Ober {et~al.}(2015)Ober, Wolf, Uribe, \& Klahr}]{ober_tracing_2015}
Ober, F., Wolf, S., Uribe, A.~L., \& Klahr, H.~H. 2015, Astronomy and
  Astrophysics, 579, A105

\bibitem[{Pudritz {et~al.}(2014)Pudritz, Klassen, Kirk, Seifried, \&
  Banerjee}]{pudritz_role_2014}
Pudritz, R.~E., Klassen, M., Kirk, H., Seifried, D., \& Banerjee, R. 2014, in
  '', Vol. 302, 10--20

\bibitem[{Radford(1961)}]{radford_microwave_1961}
Radford, H.~E. 1961, Physical Review, 122, 114

\bibitem[{Rees {et~al.}(1989)Rees, Durrant, \& Murphy}]{rees_stokes_1989}
Rees, D.~E., Durrant, C.~J., \& Murphy, G.~A. 1989, The Astrophysical Journal,
  339, 1093

\bibitem[{Reissl {et~al.}(2016)Reissl, Brauer, \& Wolf}]{reissl_radiative_2016}
Reissl, S., Brauer, R., \& Wolf, S. 2016, arXiv:1604.05305 [astro-ph], arXiv:
  1604.05305

\bibitem[{Reissl {et~al.}(2014)Reissl, Wolf, \& Seifried}]{reissl_tracing_2014}
Reissl, S., Wolf, S., \& Seifried, D. 2014, Astronomy and Astrophysics, 566,
  A65

\bibitem[{Roberts {et~al.}(1995)Roberts, Crutcher, \&
  Troland}]{roberts_distribution_1995}
Roberts, D.~A., Crutcher, R.~M., \& Troland, T.~H. 1995, The Astrophysical
  Journal, 442, 208

\bibitem[{Schadee(1978)}]{schadee_zeeman_1978}
Schadee, A. 1978, Journal of Quantitative Spectroscopy and Radiative Transfer,
  19, 517

\bibitem[{Schöier {et~al.}(2005)Schöier, van~der Tak, van Dishoeck, \&
  Black}]{schoier_atomic_2005}
Schöier, F.~L., van~der Tak, F. F.~S., van Dishoeck, E.~F., \& Black, J.~H.
  2005, Astronomy and Astrophysics, 432, 369

\bibitem[{Seifried \& Walch(2015)}]{seifried_impact_2015}
Seifried, D. \& Walch, S. 2015, Monthly Notices of the Royal Astronomical
  Society, 452, 2410

\bibitem[{Troland \& Crutcher(2008)}]{troland_magnetic_2008}
Troland, T.~H. \& Crutcher, R.~M. 2008, The Astrophysical Journal, 680, 457

\bibitem[{Wells(1999)}]{wells_rapid_1999}
Wells, R.~J. 1999, Journal of Quantitative Spectroscopy and Radiative Transfer,
  62, 29

\bibitem[{Wiles {et~al.}(2016)Wiles, Lo, Redman, Cunningham, Jones, Burton, \&
  Bronfman}]{wiles_scaled_2016}
Wiles, B., Lo, N., Redman, M.~P., {et~al.} 2016, Monthly Notices of the Royal
  Astronomical Society, 458, 3429

\bibitem[{Wilson {et~al.}(1997)Wilson, Walker, \&
  Thornley}]{wilson_density_1997}
Wilson, C.~D., Walker, C.~E., \& Thornley, M.~D. 1997, The Astrophysical
  Journal, 483, 210

\bibitem[{Wolf {et~al.}(2003)Wolf, Launhardt, \& Henning}]{wolf_magnetic_2003}
Wolf, S., Launhardt, R., \& Henning, T. 2003, The Astrophysical Journal, 592,
  233

\end{thebibliography}

\begin{appendix}
 \section{Derivation of the analysis method}\label{appendix_1}
In the following, the Eq. \ref{eq:mag_field} which is used to analyze Zeeman split spectral lines is derived (see Sect. \ref{analysis_method}). The first step is to calculate the derivative of the intensity $I$ with respect to the frequency $\nu$:
  \begin{align}
   I=&F(\nu_0+\Delta\nu_z-\nu,a)\cdot(1+\cos^2\theta) 
   + 2F(\nu_0-\nu,a)\cdot\sin^2\theta\nonumber\\
   + &F(\nu_0-\Delta\nu_z-\nu,a)\cdot(1+\cos^2\theta),\\
   \frac{\mathrm{d}I}{\mathrm{d}\nu}=&
   \frac{\mathrm{d}F(\nu_0+\Delta\nu_z-\nu,a)}{\mathrm{d}\nu}\cdot(1+\cos^2\theta) 
   + 2\frac{\mathrm{d}F(\nu_0-\nu,a)}{\mathrm{d}\nu}\cdot\sin^2\theta\nonumber\\
   + &\frac{\mathrm{d}F(\nu_0-\Delta\nu_z-\nu,a)}{\mathrm{d}\nu}\cdot(1+\cos^2\theta).\label{eq:derived_F}
  \end{align}
  If we assume $\Delta\nu_z\rightarrow0$, the line function can be written as a Taylor series:
  \begin{align}
   F(\nu_0+\Delta\nu_z-\nu,a) \approx F(\nu_0-\nu,a) + \frac{\mathrm{d}F(\nu_0-\nu,a)}{\mathrm{d}\nu} \cdot \Delta\nu_z,\\
   \frac{\mathrm{d}F(\nu_0-\nu,a)}{\mathrm{d}\nu} \cdot \Delta\nu_z \approx F(\nu_0+\Delta\nu_z-\nu,a) - F(\nu_0-\nu,a).
  \end{align}
  In combination with the following assumptions, we obtain each derivative of the line shape in Eq. \ref{eq:derived_F}:
  \begin{align}
   \frac{\mathrm{d}F(\nu_0+\Delta\nu_z-\nu,a)}{\mathrm{d}\nu} \cdot \Delta\nu_z \approx& F(\nu_0+\Delta\nu_z-\nu,a) \nonumber\\
   -& F(\nu_0-\nu,a), \label{eq:assumption_1}\\
   \frac{\mathrm{d}F(\nu_0-\Delta\nu_z-\nu,a)}{\mathrm{d}\nu} \cdot \left(-\Delta\nu_z\right) \approx& F(\nu_0-\Delta\nu_z-\nu,a) \nonumber\\
   -& F(\nu_0-\nu,a), \label{eq:assumption_2}\\
   2\frac{\mathrm{d}F(\nu_0-\nu,a)}{\mathrm{d}\nu} \cdot \Delta\nu_z \approx& \frac{\mathrm{d}F(\nu_0+\Delta\nu_z-\nu,a)}{\mathrm{d}\nu}\cdot \Delta\nu_z \nonumber\\
   +& \frac{\mathrm{d}F(\nu_0-\Delta\nu_z-\nu,a)}{\mathrm{d}\nu}\cdot \left(-\Delta\nu_z\right). \label{eq:assumption_3}
  \end{align}
  Finally, we obtain the equation for the Zeeman analysis (Sect. \ref{analysis_method}, Eq. \ref{eq:mag_field}) by using Eqs. \ref{eq:assumption_1} - \ref{eq:assumption_3} and multiplying Eq. \ref{eq:derived_F} with $\Delta\nu_z\cos\theta$:
  \begin{align}
   \left(\frac{\mathrm{d}I}{\mathrm{d}\nu}\right)\Delta\nu_z\cos\theta =& 2F(\nu_0+\Delta\nu_z-\nu,a)\cos\theta \nonumber\\
   -& 2F(\nu_0-\Delta\nu_z-\nu,a)\cos\theta,\\
   \left(\frac{\mathrm{d}I}{\mathrm{d}\nu}\right)\Delta\nu_z\cos\theta =& V.
  \end{align}
 
\end{appendix}

\end{document}